\documentclass[12pt]{article}
\usepackage{amsfonts,amsmath,amssymb}
\usepackage{graphicx}
\usepackage{epic,eepic,color,graphicx}

\newfont{\twelvemsb}{msbm10 scaled\magstep1}
\newfont{\eightmsb}{msbm8}
\newfam\msbfam
\textfont\msbfam=\twelvemsb \scriptfont\msbfam=\eightmsb
\catcode`\@=11
\def\Bbb{\ifmmode\let\next\Bbb@\else
\def\next{\errmessage{Use \string\Bbb\space only in math mode}}\fi\next}
\def\Bbb@#1{{\fam\msbfam{{#1}}}}

\newcommand{\be}{\begin{equation}}
\newcommand{\ee}{\end{equation}}
\newcommand{\ba}{\begin{eqnarray}}
\newcommand{\ea}{\end{eqnarray}}

\begin{document}

\sloppy
\renewcommand{\thefootnote}{\fnsymbol{footnote}}
\newpage
\setcounter{page}{1} \vspace{0.7cm}
\begin{flushright}
20/01/09
\end{flushright}
\vspace*{1cm}
\begin{center}
{\bf On the high spin expansion in the $sl(2)$ ${\cal N}=4$ SYM theory}\\
\vspace{1.8cm} {Davide Fioravanti $^a$, Gabriele Infusino $^b$,
Marco Rossi $^b$}\\
\vspace{.5cm} $^a$ {\em Sezione INFN di Bologna, Dipartimento di Fisica,
Universit\`a  di Bologna} \\
{\em Via Irnerio 46, Bologna, Italy}\\
\vspace{.3cm} $^b${\em Dipartimento di Fisica dell'Universit\`a della Calabria
and INFN, Gruppo collegato di Cosenza,} \\
{\em I-87036 Arcavacata di Rende, Cosenza, Italy}
\end{center}
\renewcommand{\thefootnote}{\arabic{footnote}}
\setcounter{footnote}{0}
\begin{abstract}
{\noindent We study the the high spin expansion of the anomalous dimension for long operators belonging to  the
$sl(2)$ sector of ${\cal N}=4$ SYM. Keeping the ratio $j$ between
the twist and the logarithm of the spin fixed, the anomalous
dimensions expand as $\gamma= f(g,j)\ln s + f^{(0)}(g,j)+O(1/\ln s)$.
This particular double scaling limit is efficiently described, up to the
desired accuracy $O(\ln s ^0 )$, in terms of linear integral equations.
By using them, we are able to evaluate both at weak and strong
coupling the sub-leading scaling function $f^{(0)}(g,j)$ as series in $j$, up to the
order $j^5$. Thanks to these results, the possible extension of the liaison with the $O(6)$ non-linear sigma model may be tackled on a solid ground.}
\end{abstract}
\vspace{6cm}

\newpage

\section{Introduction}
\setcounter{equation}{0}

The set of composite operators
\begin{equation}
{\mbox {Tr}}({\cal D}^s{\cal Z}^L)+\ldots \, , \label {op}
\end{equation}
where ${\cal D}$ is a covariant derivative acting in all possible ways
on the complex scalar field ${\cal Z}$, constitutes the so-called $sl(2)$ sector of
${\cal N}=4$ SYM theory. The integer numbers $s$ and $L$ are
called spin and twist, respectively. In the framework of the
AdS-CFT correspondence \cite {MWGKP} this set of operators
received particular attention, also because of its connection
\cite {LIP,BDM} to twist operators in QCD. In the high spin limit
anomalous dimensions $\gamma (g,L,s)$ of (\ref {op}) show a
logarithmic divergence,
\begin{equation}
\gamma (g,L,s)= \ln s \ f(g) + f_c(g,L) + o(s^0) \, . \label {lns}
\end{equation}
The function of the coupling $f(g)$, which equals twice the cusp
anomalous dimension of light-like Wilson loops, is known also as
universal scaling function, where the term 'universal' takes into
account its lack of dependance on the (fixed) twist $L$. This
property, in particular, must imply that $f(g)$ is free from
wrapping effects, which makes the asymptotic Bethe Ansatz
equations \cite {MZ,BS,AFS,BES} predictive for its exact
determination. On the other hand, the function $f_c(g,L)$ is not
'universal', but, since it is related through functional relations
\cite {BKDM,BFTT} to $f(g)$, it contains a universal part.
Finally, one has to mention that both $f$ and $f_c$ are related
\cite {DMS} to form factors in scattering amplitudes. The function
$f(g)$ can be found from the solution of a linear
integral equation \cite {ES,BES} derived from asymptotic Bethe
Ansatz. This equation was solved both at the weak \cite {ES,BES} and,
more importantly, at the strong coupling \cite {BBKS,BKK,KSV} limit.
Moreover, the function $f_c(g,L)$ was studied in \cite
{FRS} (weak coupling), \cite {BFTT} (strong coupling from string
theory) and in the contemporaneous paper \cite {FZ,FGR4} (strong
coupling, using linear integral equations derived from gauge
theory Bethe Ansatz).

\medskip

The structure (\ref {lns}) of the high spin anomalous dimension is
preserved if one introduces the following scaling limit,
\begin{equation}
s \rightarrow +\infty \, , \quad L \rightarrow +\infty \, , \quad
j=\frac {L-2}{\ln s} \, \, {\mbox {fixed}} \, , \label {jlimit}
\end{equation}
proposed\footnote {To be precise the limit considered in \cite
{AM,FRS} is  \begin{equation} s \rightarrow +\infty \, , \quad L
\rightarrow +\infty \, , \quad j=\frac {L}{\ln s} \, \, {\mbox
{fixed}} \, . \nonumber
\end{equation} Referring to expansion (\ref {lnsinfty}),
the different limit (\ref {jlimit}) does not affect $f(g,j)$, but
gives easier forms for $f^{(0)}(g,j)$.} in \cite {AM} for the
strong coupling (string theory) case and then in \cite {FRS} for
the weak coupling ${\cal N}=4$ SYM theory. Similar limits were
also studied in one loop ${\cal N}=4$ SYM - in \cite {BGK}, when
$\frac {L}{\ln (s/L)}$ is fixed  - and in the strong coupling
${\cal N}=4$ SYM \cite {AFS,CK,GRO,V} and string \cite {BGK,FTT,RT} cases,
when $\frac {g L}{\ln (s/L)}$ is fixed.

As we will show, in the case of limit (\ref {jlimit}), relation (\ref {lns}) is
replaced by
\begin{equation}
\gamma (g,j,s)=  f(g,j) \ln s + f^{(0)}(g,j) + \sum
_{k=1}^{\infty} {f^{(k)}(g,j) (\ln s)}^{-k} + O({(\ln s)}
^{-\infty})
 \, , \label {lnsinfty}
\end{equation}
where the notation $O({(\ln s)} ^{-\infty})$ stands for terms
going to zero faster than any powers $(\ln s)^{-k}$, $k \in
{\mathbb N}$. Importantly, we may conjecture that wrapping effects can affect none of
the functions appearing in (\ref {lnsinfty}), which thus are still completely determined by the asymptotic
Bethe Ansatz equations or (non-)linear integral equation \cite{FGR4}. One reason for that is the matching comparison with the string expansion of \cite{BFTT}. 

At small $j$ the structure of the function $f(g,j)$, called
often generalised scaling function,
\begin{equation}
f(g,j)=\sum _{n=0}^{\infty }f_n(g) j^n \, ,
\end{equation}
was investigated at weak coupling \cite {FRS} and at strong
coupling \cite {FGR1,BK,FGR2,FGR3}. The generalised scaling function $f(g,j)$
for $j \ll g$ was also shown \cite {BK} to coincide with the
energy density \cite {BF} of the nonlinear $O(6)$ sigma model
embedded into $AdS_5 \times S^5$, thus confirming the related previous
proposal \cite {AM} by Alday and Maldacena. Results for $j \gg g$
are also available \cite {BEC}.

The aim of this paper is to investigate the structure of
$f^{(0)}(g,j)$ at very small $j$, both at the weak and the strong
coupling limit. We expect $f^{(0)}(g,j)$ to be an analytical
function of $j$, so that the expansion
\begin{equation}
f^{(0)}(g,j)=\sum _{n=0}^{\infty }f_n^{(0)}(g) j^n \,  \label
{f0n}
\end{equation}
will capture all its properties. Even if we report explicit
results on functions $f_n^{(0)}(g)$ for $n \leq 5$, we will show
that in general the various $f_n^{(0)}(g)$ can be obtained using a
linear integral equation - giving the density of Bethe roots and
of the so-called holes - which is equivalent to the asymptotic
Bethe Ansatz equations in the high spin limit if one neglects
terms of order $ (\ln s )^{-1} $. This equation follows from
previous results, mainly contained in \cite {BFR1,BFR2}, and
generalises the one we used previously \cite {FGR1,FGR2,FGR3} to
determine the components $f_n(g)$ of the scaling function $f(g,j)$
and which is equivalent to the 'FRS' equation, proposed in \cite {FRS}.

The plan of this paper is as follows.

In Section 2 we report on the one loop results. We show that the
density of Bethe roots and holes can be found by solving a linear
integral equation, which is exact if one neglects terms of order $O(1/\ln
s)$ and we give the expressions for $f_n^{(0)}(g=0)$, when $n=1, \ldots
,5$.

In Section 3 we write the linear integral equation satisfied by
the density of roots and holes at arbitrary values of the coupling
constant. Using such density, we perform weak coupling
computations and give $f_n^{(0)}(g)$ up to three loops and up
to $n=4$.

In Section 4 we re-write the linear integral equation for the
density as a set of linear systems: from the solution to the
$n$-th system one can get the $n$-th component in the expansion
(\ref {f0n}). The solution to the $n$-th system depends - very
similarly to the systems describing the components $f_n(g)$ of the
generalised scaling function - on the previous ones. This
reformulation of the problem simplifies the analysis of the strong
coupling limit, which we do up to $n=5$.

\section{One loop results}
\setcounter{equation}{0}

We begin our study by considering the $sl(2)$ sector of ${\cal
N}=4$ SYM theory at one loop. In order to distinguish them from the
exact (i.e. all loops) correspondents, all the one loop quantities (such as
the density $\sigma$ and the roots-holes separator $c$) will
be denoted by an index $0$. As stated in the Introduction, we
are interested in the limit (\ref {jlimit})
\begin{equation}
s \rightarrow +\infty \, , \quad L \rightarrow +\infty \, , \quad
j=\frac {L-2}{\ln s} \, \, {\mbox {fixed}} \, . \nonumber
\end{equation}
In this limit, the density of Bethe roots and holes at one loop in
perturbation theory, $\sigma _0(u)$, satisfies the linear equation
\begin{eqnarray}
\sigma _0 (u)&=&L \left [ \psi \left (\frac {1}{2}-iu \right )+
 \psi \left (\frac {1}{2}+iu \right ) \right ]- \psi \left (1-iu -i\frac {s}{{\sqrt {2}}}\right )-
 \psi \left (1-iu +i\frac {s}{{\sqrt {2}}}\right ) -  \nonumber \\
 &-& 
\psi \left (1+iu -i\frac {s}{{\sqrt {2}}}\right )- \psi \left
(1+iu +i\frac {s}{{\sqrt {2}}}\right ) - 2\ln 2 + \nonumber \\
&+& \int _{-c_0}^{c_0}\frac {dv}{2\pi} \left [ \psi (1-iu+iv)+
\psi (1+iu+iv) \right ] \sigma _0(v) + O\left (\frac {1}{\ln s}
\right ) \, ,\label {S0u}
\end{eqnarray}
which, as we wrote, is exact if we neglect terms of order $O(1/\ln
s)$. Further, we have to impose the condition
\begin{equation}
 \int _{-c_0}^{c_0}du \sigma _0 (u) = -2\pi \, j \, \ln s  +
O\left (\frac {1}{\ln s} \right ) \, , \label {s0condu}
\end{equation}
which comes from the fact that the parameter $c_0$ defines the interval
$[-c_0,c_0]$ in which the $L-2$ internal holes concentrate,

It is convenient to use the Fourier transform, $\hat \sigma _0(k)$, which
satisfies the linear equation
\begin{eqnarray}
\hat \sigma _0(k)&=&-2\pi \frac {\frac {L}{2}\left (1-e^{-\frac {|k|}{2}}
\right )+e^{-\frac {|k|}{2}}\left (1-
\cos \frac {ks}{\sqrt {2}}\right )}
{\sinh \frac {|k|}{2}}-4\pi \delta (k) \ln 2 - \nonumber \\
&-& \frac {e^{-\frac {|k|}{2}}}{\sinh \frac
{|k|}{2}} \int  _{-\infty}^{+\infty} \frac {dh}{2\pi}
\hat \sigma _0(h) \left [ \frac {\sin (k-h)c_0}{k-h}-
\frac {\sin hc_0}{h}\right ] +
O\left (\frac {1}{\ln s} \right )
 \, , \label {s0}
\end{eqnarray}
and the condition
\begin{equation}
2 \int  _{-\infty }^{+\infty} \frac {dk}{2\pi}
\hat \sigma _0(k) \frac {\sin kc_0}{k} = -2\pi \, j \, \ln s  +
O\left (\frac {1}{\ln s} \right ) \, . \label {s0cond}
\end{equation}
We can now give justifications of these formul{\ae}. We start from
(3.52) of \cite {BFR2}, where the $L-2$ in the last term of the
first line is replaced by a sum on the internal holes $\sum
\limits _{h=1}^{L-2}e^{iku_h}$. This equation, for $\frac
{d}{du}F_0(u)=\sigma _0(u)+O(1/s)$, is exact if we neglect terms
of order $O(1/s)$. Now, in order to express the sum on internal
holes in terms of the density $\sigma _0(u)$, we use results from
Appendix A. Referring to formula (\ref {sumh}), we realize that,
if one neglects terms of order $O(1/\ln s)$, such sum is given by
the first term in the right hand side of (\ref {sumh}), which is
linear in the density $\sigma _0(u)=\frac {d}{du}Z_0(u)$. After
doing this, we get expression (\ref {s0}) for the equation
satisfied by the density of Bethe roots and holes in the Fourier
space \footnote {Actually, in order to get equation (\ref {s0}), we
have to use once identity (\ref {s0cond}).} . In a completely
analogous way, when we express the counting of internal holes in
terms of $\sigma _0(u)$, we get (\ref {s0condu}, \ref {s0cond}).

As a further verification of the correctness of our starting
equations, in the first part of Appendix B we prove that equation
(\ref {S0u}) is compatible with the corresponding equation - valid
for very small $u$ -  that can be deduced from the results of
\cite {BGK}.

Having established on quite firm ground our starting point, we
pass to analyze in detail the behaviour of the various quantities
in the limit (\ref {jlimit}). Consistency considerations coming
from the analysis of (\ref {s0}, \ref {s0cond}) imply that in such
limit the parameter $c_0$, depending, via (\ref {s0}, \ref
{s0cond}), on $\ln s$ and $j$, expands, when $j\ll 1$, as
\begin{equation}
c_0= \sum _{n=1}^{\infty}c_0^{(0,n)}j^n + \frac {1}{(\ln s)}
\sum _{n=1}^{\infty}c_0^{(1,n)}j^n
 + O\left (\frac {1}{(\ln s)^2} \right ) \, . \label {c0exp}
\end{equation}
Expanding also the condition (\ref {s0cond}) we have
\begin{equation}
2 \sigma _0 (0) c_0 + \frac {1}{3} {\sigma }^{''}_0(0) c_0 ^3 + \ldots
= -2\pi j \ln s \, ,  \label {s0cond1}
\end{equation}
where we use the notation $\sigma _0(0)$, ${\sigma }^{''}_0 (0)$ to indicate
the values at $u=0$ of the function $\sigma _0$ and its second
derivative, respectively.
For such quantities expansions similar to (\ref {c0exp}) hold, e.g.
\begin{equation}
\sigma _0(0)=\ln s [\sum _{n=0}^{\infty}\sigma_0^{(-1,n)}(0)j^n]+
[\sum _{n=0}^{\infty}\sigma_0^{(0,n)}(0)j^n]+
O\left (\frac {1}{\ln s} \right ) \, . \label {sigma0exp}
\end{equation}
Explicitly we have
\begin{equation}
\sigma _0 (0)=[-4 - 4j \ln 2+O(j^3)]\ln s - [8 \ln 2 + 4 \gamma _E+O(j^3)]  +
O\left (\frac {1}{\ln s} \right ) \label {sigma0}
\end{equation}
and also that
\begin{equation}
{\sigma }^{''}_0(0)= [56 \zeta (3) + O(j)]+O\left (\frac {1}{\ln s} \right )
 \label {sigma20} \, .
\end{equation}
Now, inserting the expansion (\ref {c0exp}) in the condition (\ref {s0cond1})
and using also (\ref {sigma0}) we can start finding, by equating equal powers in $j$ and $\ln s$, the various coefficients $c_0^{(0,n)}$, $c_0^{(1,n)}$ .

We get, without much ado
\begin{eqnarray}
&& c^{(0,1)}= \frac {\pi}{4} \, , \quad c^{(0,2)}= - \frac {\pi}{4} \ln 2 \, ,
\quad c^{(0,3)}= \frac {\pi}{4} (\ln 2)^2 \, ;
\quad  c^{(1,1)}= - \frac {\pi}{4} (2\ln 2 + \gamma _E) \\
&&
 c^{(1,2)}= \frac {\pi}{2}\ln 2  (2\ln 2 + \gamma _E) \, , \quad
c^{(1,3)}= -\frac {3}{4}\pi (\ln 2)^2  (2\ln 2 + \gamma _E)+\frac {7}{192}\pi ^3 \zeta (3) \, ,
\end{eqnarray}
in such a way that one can write
\begin{eqnarray}
c_0&=&\left [ \frac {\pi}{4} {j}- \frac {\pi}{4} \ln 2 \ j^2
+ \frac {\pi}{4} (\ln 2)^2 j^3 + O(j^4) \right ] + \Bigl [
- \frac {\pi}{4} (2\ln 2 + \gamma _E) j
 + \label {c0} \\
&+& \frac {\pi}{2}\ln 2  (2\ln 2 + \gamma _E) j^2 -\left ( \frac {3}{4}\pi (\ln 2)^2  (2\ln 2 + \gamma _E)-\frac {7}{192}\pi ^3 \zeta (3) \right ) j^3
+ O(j^4) \Bigr ] \frac {1}{\ln s}  +
O\left (\frac {1}{(\ln s)^2} \right ) \, . \nonumber
\end{eqnarray}
All this information can be used to compute the anomalous
dimension at one loop $\gamma _{g^2}$ up to the desired order in
$j$. From the formul{\ae} - exact if we neglect terms $O\left
(\frac {1}{\ln s} \right )$ -
\begin{eqnarray}
&&\gamma _{g^2}=g^2 E_0 \, , \quad E_0=- \int _{-\infty}^{\infty}
\frac {du}{2\pi} e(u) \sigma  _0 (u)+
\int _{-c_0}^{c_0} \frac {du}{2\pi} e(u) \sigma  _0 (u) = \\
&=& - \int _{-\infty}^{\infty} \frac {dk}{4\pi^2} \hat e(k) \hat
\sigma  _0 (k)+ \int _{-\infty}^{\infty} \frac {dk}{4\pi^2} \hat e(k)
\int _{-\infty}^{\infty} \frac {dh}{2\pi} \hat \sigma  _0 (h)
\left [ 2\frac {\sin (k-h)c_0}{k-h}-
2 \frac {\sin hc_0}{h}\right ] - \ln s j e(0) \, , \nonumber
\end{eqnarray}
where $E_0$ denotes the energy of the spin $-1/2$ Heisenberg chain
and the function
\begin{equation}
e(u)=\frac {1}{u^2 +\frac {1}{4}} \Rightarrow \hat e (k)=2 \pi e^{-\frac {|k|}{2}} \, ,
\end{equation}
we get, expanding in powers of $j$:
\begin{eqnarray}
\gamma _{g^2}&=& g^2 \ln s \Bigl [ 4 - 4 \ln 2 \, j+\frac {7 \zeta
(3) \pi ^2}{24} \, j^3 -
\frac {7 \zeta (3) \pi ^2 \ln 2}{12} \, j^4 + \left (\frac{7\pi^2 (\ln 2)^2 \zeta(3)}{8}-\frac{31\pi^4\zeta(5)}{640}\right) j^5 + O(j^6) \Bigr ] + \nonumber \\
&+& g^2 \Bigl [ 4 \gamma _E - \frac {7 \zeta (3) \pi ^2}{12} (2\ln
2 +\gamma _E) \, j^3+ \frac {7 \zeta (3) \pi ^2 \ln 2}{4} (2\ln 2
+\gamma _E) j^4 -
\nonumber \\
&-&  \left (\frac{7\pi^2 (\ln 2)^2 \zeta(3)}{2}-\frac{31\pi^4\zeta(5)}{160}\right) (2\ln 2 + \gamma _E) j^5 +  \frac{49\pi^4 \zeta^2 (3)}{960} j^5
+ O(j^6) \Bigr ] +
O\left (\frac {1}{\ln s} \right ) \label {E0} \, .
\end{eqnarray}
As we show in Appendix B, one can also get such an expression
improving the results of \cite {BGK} in order to include also
contributions of order $(\ln s)^0 j^k$ (as far as we know, authors
of \cite {BGK} are interested in order $\ln s j^k$ contributions
and develop their calculations accordingly). This reinforce our
belief in the goodness of (\ref {S0u}) as description of the
leading and subleading high spin behaviour in one loop $sl(2)$
sector of ${\cal N}=4$ SYM.

\section{All loops equations}
\setcounter{equation}{0}

Let us now pass to study the anomalous dimension as a function of
the coupling constant $g$. As usual, we split the density of roots
and holes, $\sigma (u)$, as $\sigma (u)=\sigma _0(u)+
\sigma_H(u)$, where $\sigma _0(u)$ is the one loop contribution
and $\sigma _H(u)$ is the higher than one loop contribution to the
actual density. It is convenient to introduce the quantity $S(k)$,
related to the Fourier transforms of $\sigma _0(u)$ and $\sigma
_H(u)$ as
\begin{equation}
 S(k)=\frac {2\sinh \frac {|k|}{2}}{2\pi |k|}\left [\hat \sigma _H (k)
-\frac {e^{-\frac {|k|}{2}}}{\sinh \frac {|k|}{2}}
\int  _{-\infty }^{+\infty} \frac {dp}{2\pi}
\hat \sigma _0(p)
\frac {\sin (k-p)c_0} {k-p}+
\frac {e^{-\frac {|k|}{2}}}{\sinh \frac {|k|}{2}}
\int  _{-\infty }^{+\infty} \frac {dp}{2\pi}
\hat \sigma (p)
\frac {\sin (k-p)c} {k-p} \right ] \, , \label{Sndefi}
\end{equation}
where $c$ indicates the separator between the holes
- concentrated in the interval $[-c,c]$ - and the Bethe roots. 
  
The function $S(k)$ satisfies the linear equation
\begin{eqnarray}
&&S(k)=\frac {L}{|k|}[1-J_0({\sqrt {2}}gk)]
+  \frac {1}{\pi {|k|}} \int _{-\infty }^{+\infty}
\frac {dh}{|h|} \Bigl [ \sum _{r=1}^{\infty}
r (-1)^{r+1}J_r({\sqrt {2}}gk) J_r({\sqrt {2}}gh)\frac {1-{\mbox {sgn}}(kh)}{2}
e^{-\frac {|h|}{2}} + \nonumber \\
&+&{\mbox {sgn}} (h) \sum _{r=2}^{\infty}\sum _{\nu =0}^{\infty} c_{r,r+1+2\nu}(-1)^{r+\nu}e^{-\frac {|h|}{2}} \Bigl (
J_{r-1}({\sqrt {2}}gk) J_{r+2\nu}({\sqrt {2}}gh)-
J_{r-1}({\sqrt {2}}gh)
J_{r+2\nu}({\sqrt {2}}gk)\Bigr ) \Bigr ] \cdot \nonumber \\
&\cdot& \Bigl \{ \frac {\pi |h|}{\sinh \frac {|h|}{2}}S(h)-4\pi \ln 2 \ \delta (h)-\pi j
\ln s \frac {1-e^{\frac {|h|}{2}}}{\sinh \frac {|h|}{2}}-2\pi \frac {1-
e^{-\frac {|h|}{2}}\cos \frac {hs}{{\sqrt {2}}}}{\sinh \frac {|h|}{2}}
- \nonumber \\
&-& \frac {e^{\frac {|h|}{2}}}{\sinh \frac {|h|}{2}}\int  _{-\infty }^{+\infty} \frac {dp}{2\pi} \hat \sigma (p) \left [ \frac {\sin (h-p)c} {h-p}-\frac {\sin pc}{p}\right ] \Bigr \} +
O\left (\frac {1}{\ln s} \right ) \, . \label{Skeq}
\end{eqnarray}
And this relation has to be solved together with the condition
\begin{equation}
 2 \int  _{-\infty }^{+\infty} \frac {dk}{2\pi}
\hat \sigma (k) \frac {\sin kc}{k} = -2\pi j \ln s +
O\left (\frac {1}{\ln s} \right )  \, . \label {scond}
\end{equation}
To justify equation (\ref {Skeq}) we start from (4.11) of \cite
{BFR2}, where in the last term of the first line $L-2$ is replaced
by a sum on the internal holes $\sum \limits _{h=1}^{L-2}e^{iku_h}$ and
the $0$ in the arguments by $u_h$. Moreover, in the integral in
the second line we can replace the extremes $\pm b_0$ with $\pm
\infty$ and $\frac {d}{dv} F_0(v)$ with $\sigma _0(v)$ (\ref
{S0u}). The equation obtained in such a way for $\frac {d}{dv}
F^H(v)=\sigma _H (v)+O(1/s)$ is exact when neglecting terms of
order $O(1/s)$. Now, we can use results in Appendix A \footnote
{We remember that $\sigma (u)=\frac {d}{du}Z(u)$, where $Z(u)$ is
the counting function.}, which are valid for a general counting
function $Z(u)$ such that $Z(c)=-\pi (L-2) + O(1/\ln s)$, which is
exactly condition (\ref {scond}). Using (\ref {sumh}), we evaluate
the sum over internal holes keeping only the first (linear) term
in the right hand side of this formula, since we want to neglect
$O(1/\ln s)$ contributions. After passing to Fourier transforms
and defining (\ref {Sndefi}), we get eventually equation (\ref
{Skeq}).

Finally, we notice that the relation of the function $S(k)$ with
the anomalous dimension $\gamma (g)$ is a generalisation of an
identity found in \cite {KL}:
\begin{equation}
\gamma (g)=2 S(0) \, . \label {E-s}
\end{equation}
The proof of such identity follows by the simple comparison
between (\ref {Skeq}) and the expression for the anomalous
dimension
\begin{eqnarray}
\gamma (g)&=&- \int _{-\infty}^{\infty} \frac {du}{2\pi} q_2(u)
\sigma (u)+ \int _{-c}^{c} \frac {du}{2\pi} q_2(u) \sigma   (u) \,
, \quad q_2(u)=\frac {i}{x^+(u)}-\frac {i}{x^-(u)} \, , \nonumber
\\
x^{\pm}(u)&=&\frac {u\pm 1/2}{2}\left [1+\sqrt {1-\frac
{2g^2}{(u\pm 1/2)^2 }}\right ] \, . \label {E-high}
\end{eqnarray}
A little care is needed in order to interpret (\ref {E-s}): the
term proportional to $g^{2n}$, which depends on the density at
$n+1$ loops, gives the $n$-th loop contribution to the anomalous
dimension.

In the next subsection, we will compute at two and three loops the
anomalous dimension in the limit (\ref {jlimit}). New results will
be the subleading, $O(\ln s ^0)$ contribution, which we will
evaluate as power series in $j$, up to $j^4$.

\subsection{Two loops anomalous dimension}

Let us expand the linear equation for $S(k)$ at order $g^4$, i.e.
at three loops approximation for the density. Before doing this,
it is convenient to write the expression for $S(k)$ at the order
$g^2$. From the general expression we get
\begin{eqnarray}
S_{g^2}(k)&=&\frac {g^2}{2} L |k| + \frac {1}{\pi |k|} \int _{-\infty }^{+\infty}
\frac {dh}{|h|} \frac {{\sqrt {2}}gh}{2}\frac {\sqrt {2}gk}{2}\frac {1-{\mbox {sgn}}(kh)}{2}e^{-\frac {|h|}{2}} \Bigl \{ -4\pi \ln 2 \ \delta (h)
- \label {Sk2} \\
&-& \pi j
\ln s \frac {1-e^{\frac {|h|}{2}}}{\sinh \frac {|h|}{2}}-2\pi \frac {1-
e^{-\frac {|h|}{2}}\cos \frac {hs}{{\sqrt {2}}}}{\sinh \frac {|h|}{2}}
- \frac {e^{\frac {|h|}{2}}}{\sinh \frac {|h|}{2}}\int  _{-\infty }^{+\infty} \frac {dp}{2\pi} \hat \sigma _0(p) \left [ \frac {\sin (h-p)c_0} {h-p}-\frac {\sin pc_0}{p}\right ] \Bigr \} \nonumber \, .
\end{eqnarray}
Comparing (\ref {E-s}) with (\ref {Sk2}) we get
\begin{equation}
S_{g^2}(k)=\frac {g^2}{2} L |k|+ \frac {1}{2} \gamma _{g^2} \, .
\label {Sk2-E1}
\end{equation}
Let us pass now to the three loops (order $g^4$) density. In this
case the relevant equation reads
\begin{eqnarray}
S_{g^4}(k)&=&-\frac {g^4}{16} L |k|^3 + \frac {1}{\pi |k|} \int _{-\infty }^{+\infty}
\frac {dh}{|h|} \frac {{\sqrt {2}}gh}{2}\frac {\sqrt {2}gk}{2}\frac {1-{\mbox {sgn}}(kh)}{2}e^{-\frac {|h|}{2}} \Bigl \{ \frac {\pi |h|}{\sinh \frac {|h|}{2}}
S_{g^2}(h)- \nonumber \\
&-&  \frac {e^{\frac {|h|}{2}}}{\sinh \frac {|h|}{2}}\int  _{-\infty }^{+\infty} \frac {dp}{2\pi} \hat \sigma (p) \left [ \frac {\sin (h-p)c} {h-p}-\frac {\sin pc}{p}\right ]_{g^2} \Bigr \}+ \frac {1}{\pi |k|} \int _{-\infty }^{+\infty}
\frac {dh}{|h|} \Bigl [ -\frac {1}{8} g^4 k^2 h^2 - \nonumber \\
&-& \frac {({\sqrt {2}}gh)^3}
{16}\frac {\sqrt {2}gk}{2} - \frac {({\sqrt {2}}gk)^3}
{16}\frac {\sqrt {2}gh}{2}\Bigr ] \frac {1-{\mbox {sgn}}(kh)}{2}e^{-\frac {|h|}{2}} \Bigl \{ -4\pi \ln 2 \ \delta (h)
- \label {Sk4} \\
&-& \pi j
\ln s \frac {1-e^{\frac {|h|}{2}}}{\sinh \frac {|h|}{2}}-2\pi \frac {1-
e^{-\frac {|h|}{2}}\cos \frac {hs}{{\sqrt {2}}}}{\sinh \frac {|h|}{2}}
- \frac {e^{\frac {|h|}{2}}}{\sinh \frac {|h|}{2}}\int  _{-\infty }^{+\infty} \frac {dp}{2\pi} \hat \sigma _0(p) \left [ \frac {\sin (h-p)c_0} {h-p}-\frac {\sin pc_0}{p}\right ] \Bigr \} \nonumber \, .
\end{eqnarray}
It follows that the two loops contribution to the anomalous
dimension, $\gamma _{g^4}$, is given by
\begin{eqnarray}
\gamma _{g^4}&=&-\frac {g^2}{2\pi} \int _{-\infty }^{+\infty} dh
e^{-\frac {|h|}{2}} \Bigl \{ \frac {\pi |h|}{\sinh \frac {|h|}{2}}
S_{g^2}(h)-  \frac {e^{\frac {|h|}{2}}}{\sinh \frac {|h|}{2}}\int  _{-\infty }^{+\infty} \frac {dp}{2\pi} \hat \sigma (p) \left [ \frac {\sin (h-p)c} {h-p}-\frac {\sin pc}{p}\right ]_{g^2} \Bigr \}+\nonumber \\
&+& \frac {g^4}{8\pi} \int _{-\infty }^{+\infty}
{dh} h^2  e^{-\frac {|h|}{2}} \Bigl \{ -4\pi \ln 2 \ \delta (h)
- \pi j
\ln s \frac {1-e^{\frac {|h|}{2}}}{\sinh \frac {|h|}{2}}-
 \label {E4} \\
&-&
2\pi \frac {1-
e^{-\frac {|h|}{2}}\cos \frac {hs}{{\sqrt {2}}}}{\sinh \frac {|h|}{2}}
- \frac {e^{\frac {|h|}{2}}}{\sinh \frac {|h|}{2}}\int  _{-\infty }^{+\infty} \frac {dp}{2\pi} \hat \sigma _0(p) \left [ \frac {\sin (h-p)c_0} {h-p}-\frac {\sin pc_0}{p}\right ] \Bigr \} \nonumber \, .
\end{eqnarray}
It is not difficult to compute, neglecting orders $1/s$, the
contributions to $\gamma _{g^4}$ coming from the first term in the
first line, the terms in the second line and the first term in the
third line. For these contribution we get
\begin{eqnarray}
&& \gamma _{g^4,1}=-2g^4 \zeta (3) (j\ln s +2) - \frac {g^2}{6}
\pi ^2 \gamma _{g^2} +
6g^4 \zeta (3) j\ln s-2g^4 \zeta (3) = \label {E4-1} \\
&&= -6g^4 \zeta (3) + 4g^4 \zeta (3)j\ln s -\frac {g^2}{6} \pi ^2
\gamma  _{g^2} \, . \nonumber
\end{eqnarray}
The computation of the contribution, which we indicate with
$\gamma _{g^4,2}$, coming from the last term in (\ref {E4}) relies
on the one loop results. If one wants to restrict to terms
containing powers of $j$ not higher than $4$, the relevant
integral to compute is simple:
\begin{equation}
 \gamma _{g^4,2}=\frac {g^4}{24\pi}c_0^3 \sigma _0 (0)\int _0^{+\infty} dh
\frac {h^4}
{\sinh \frac {h}{2}}=\frac {g^4}{\pi}62 \zeta (5) c_0^3 \sigma _0 (0)  \, .
\end{equation}
>From the one loop results we have, up to the desired order
\begin{eqnarray}
c_0^3 \sigma _0 (0)&=& \Bigl [ - \frac {\pi ^3}{16} j^3
+ \frac {\pi ^3}{8} j^4 \, \ln 2  \Bigr ] \ln s + \\
&+& \Bigl [  \frac {\pi ^3}{8} (2\ln 2 + \gamma _E ) j^3
-3 \frac {\pi ^3}{8} j^4 \, \ln 2 (2\ln 2 + \gamma _E ) \Bigr ] + O\left (\frac {1}{\ln s} \right ) \, ,
\end{eqnarray}
in such a way that the corresponding contribution to the energy reads
\begin{eqnarray}
 \gamma_{g^4,2}&=&\frac {g^4}{\pi} 62 \zeta (5) \Bigl [ - \frac {\pi ^3}{16} j^3
+ \frac {\pi ^3}{8} j^4 \, \ln 2  \Bigr ] \ln s + \nonumber \\
&+& \frac {g^4}{\pi} 62 \zeta (5) \Bigl [  \frac {\pi ^3}{8} (2\ln 2 +
\gamma _E ) j^3
-3 \frac {\pi ^3}{8} j^4 \, \ln 2 (2\ln 2 + \gamma _E ) \Bigr ] +
O\left (\frac {1}{\ln s} \right ) \label {E4-2} \, .
\end{eqnarray}
We are now left with the calculation of the contribution - $\gamma
_{g^4,3}$ - to the anomalous dimension coming from the last term
in the first line of (\ref {E4}). Since we restrict to terms
containing powers of $j$ not higher that $4$, at desired order
such contribution equals
\begin{equation}
 \gamma _{g^4,3}=-\frac {g^2}{6\pi}[c^3 \sigma (0)]_{g^2}\int _0^{+\infty} dh
\frac {h^2}
{\sinh \frac {h}{2}}=-g^2\frac {14}{3\pi} \zeta (3)[c^3 \sigma (0)]_{g^2} \, .
\end{equation}
We have now to compute $\sigma (0)$ and $c$ up to the order $g^2$.
For what concerns $\sigma (0)$ we have
\begin{equation}
\sigma (0)=[-4- 4j \ln 2]\ln s - [8 \ln 2 + 4 \gamma _E] + [\sigma _H(0)]_{g^2} +  O\left (\frac {1}{\ln s} \right ) \, ,
\end{equation}
where, from (\ref {Sndefi}, \ref {Sk2-E1}), one has
\begin{equation}
[\sigma _H(0)]_{g^2}= \int _{-\infty}^{\infty}
dh \frac {|h|}{2 \sinh \frac {|h|}{2}} [S(h)]_{g^2} + O(j^3) = 14 (j\ln s +2)
g^2 \zeta (3)+\frac {\pi ^2}{2} \gamma _{g^2}+O\left (\frac {1}{\ln s} \right ) \, .
\end{equation}
On the other hand, the condition (\ref {scond}) for $c$ at the desired order
reads as
\begin{equation}
2 \sigma (0) c = -2\pi j \ln s    \, .
\end{equation}
Using such relation, we can get $c$ at the order $g^2$ and up to the order
$j^2$:
\begin{eqnarray}
[c]_{g^2}&=&\left [ \frac {\pi}{16 \ln s}j(1-2\ln 2 j)+\frac {\pi}{16 (\ln s)^2} (\gamma _E+2\ln 2) (-2j +6\ln 2 j^2)\right] \cdot \nonumber \\
&\cdot & [ 28 g^2 \zeta (3)+\frac {\pi ^2}{2} \gamma _ {g^2}+ 14 j g^2 \zeta (3) \ln s ] +O\left (\frac {1}{\ln s} \right )
 \, . \label {cg2}
\end{eqnarray}
Coming back to the energy, we have
\begin{equation}
 \gamma _{g^4,3}=g^2\frac {14}{3\pi} \zeta (3)\pi j \ln s [c^2]_{g^2}+
O(j^5 , \ln s) \, ,
\end{equation}
which, after using the one loop expression for $c_0$ up to the order $j^2$,
\begin{eqnarray}
c_0&=&\left [ \frac {\pi}{4} {j}- \frac {\pi}{4} \ln 2 \ j^2
+ O(j^3) \right ] + \Bigl [
- \frac {\pi}{4} (2\ln 2 + \gamma _E) j
 + \label {c01} \\
&+& \frac {\pi}{2}\ln 2  (2\ln 2 + \gamma _E) \, j^2
+ O(j^3) \Bigr ] \frac {1}{\ln s}  +
O\left (\frac {1}{(\ln s)^2} \right ) \, . \nonumber
\end{eqnarray}
gives
\begin{eqnarray}
 \gamma _{g^4,3}&=&\frac {7}{48} \zeta (3) g^2 \pi ^2 j^3 \Bigl [ (1-3j\ln 2)+
\frac {2\ln 2 +\gamma _E}{\ln s}(-3+12\ln 2 \, j)\Bigr ] \cdot  \label {E4-3}\\
&\cdot & [ 28 g^2 \zeta (3)+\frac {\pi ^2}{2} \gamma _ {g^2}+ 14 j
g^2 \zeta (3) \ln s ] +O\left (\frac {1}{\ln s} \right ) \, .
\nonumber
\end{eqnarray}
Summing together (\ref {E4-1}, \ref {E4-2}, \ref {E4-3}) we get the energy at
the order $g^4$
\begin{eqnarray}
 \gamma _{g^4}&=& -6g^4 \zeta (3) + 4g^4 \zeta (3)j\ln s -
\frac {g^2}{6} \pi ^2 \gamma _{g^2}+ \nonumber \\
&+&\frac {g^4}{\pi} 62 \zeta (5) \Bigl [ - \frac {\pi ^3}{16} j^3
+ \frac {\pi ^3}{8} j^4 \, \ln 2  \Bigr ] \ln s + \nonumber \\
&+& \frac {g^4}{\pi} 62 \zeta (5)(2\ln 2 + \gamma _E )
\Bigl [  \frac {\pi ^3}{8}  j^3
-3 \frac {\pi ^3}{8} j^4 \, \ln 2  \Bigr ] + \label {E4tot} \\
&+& \frac {7}{48} \zeta (3) g^2 \pi ^2 j^3 \Bigl [ (1-3j\ln 2)+
\frac {2\ln 2 +\gamma _E}{\ln s}(-3+12\, j \ln 2 )\Bigr ] \cdot  \nonumber \\
&\cdot & [ 28 g^2 \zeta (3)+\frac {\pi ^2}{2} \gamma _ {g^2}+ 14 j
\, g^2 \zeta (3) \ln s ] +O\left (\frac {1}{\ln s} \right ) \, .
\nonumber
\end{eqnarray}
Alternatively, we have
\begin{eqnarray}
 \gamma _{g^4}&=&-\frac {\pi ^2}{6}g^2 \gamma _{g^2}+ g^4 \ln s \Bigl [ 4 \zeta (3) \, j -\frac {31}{8}\pi ^2 \zeta (5) \, j^3 +\frac {7}{24}\pi ^4 \zeta (3) \, j^3 + \nonumber \\
&+& \frac {31}{4}\pi ^2 \ln 2 \, \zeta (5) \, j^4 + \frac {49}{24}\pi ^2
\zeta (3)^2 \, j^4 - \frac {7}{6}\pi ^4 \zeta (3) \ln 2 \, j^4 + O(j^5) \Bigr] +\nonumber  \\
&+&g^4 \Bigl [ -6 \zeta (3)+ \frac {31}{4} \pi ^2 \zeta (5) (2\ln 2 +\gamma _E) \, j^3 + \frac {49}{12}\pi ^2 \zeta (3)^2 \, j^3 - \frac {7}{12} \pi ^4 \zeta (3) (3\ln 2 +\gamma _E) \, j^3 + \nonumber \\
&+& \frac {7}{4} \pi ^4 \zeta (3) \ln 2 (5\ln 2 +2 \gamma _E) \, j^4 -  \frac {93}{4}\pi ^2 \zeta (5) \ln 2 (2\ln 2 + \gamma _E) \, j^4 -\nonumber \\
&-& \frac {49}{8}\zeta (3)^2 \pi ^2 (4\ln 2 +\gamma _E)\, j^4 + O(j^5) \Bigr ] +O\left (\frac {1}{\ln s} \right ) \, .\label {E4tot2}
\end{eqnarray}

\subsection{Three loops anomalous dimension}

For the three loops anomalous dimension, the calculation follows the same route as in the two loops case.
We simply report the final result. The three loops anomalous dimension $\gamma _{g^6}$ expands as
\begin{equation}
\gamma _{g^6}=\ln s \sum _{n=0}^{\infty}f_{n,g^6} \, j^n +  \sum _{n=0}^{\infty}f^{(0)}_{n,g^6} \, j^n +O(1/\ln s)
\end{equation}
where the coefficients $f_{n,g^6}$ were already computed \cite {FRS} (at least, explicitly up to $n=4$) and the new coefficients $f_{n,g^6}^{(0)}$, still up to $n=4$, read as
\begin{eqnarray}
f_{0,g^6}^{(0)}&=&20 \zeta (5) + \frac {2}{3} \pi ^2 \zeta (3) + \frac {11}{45} \pi ^4 \gamma _E  \\
f_{1,g^6}^{(0)}&=&f_{2,g^6}^{(0)}=0 \\
f_{3,g^6}^{(0)}&=&-\frac {331}{1080} \pi ^6 \zeta (3) (2\ln 2 +\gamma _E)+ \frac {1}{144} \pi ^6 \zeta (3) (42\ln 2 +49\gamma _E)+ \frac {31}{3} \pi ^4 \zeta (5) (2\ln 2 +\gamma _E)- \nonumber \\
&-& \frac {31}{8} \pi ^4 \gamma _E \zeta (5)+\frac {385}{72} \pi ^4 \zeta (3)^2 - \frac {635}{8} \pi ^2 \zeta (7) (2\ln 2 +\gamma _E)
-\frac {651}{8} \pi ^2 \zeta (3) \zeta (5) \\
f_{4,g^6}^{(0)}&=&\frac {1905}{8} \pi ^2 \ln 2 (2\ln 2 +\gamma _E) \zeta (7) + \frac {1953}{16} \pi ^2 (2\ln 2 +\gamma _E) \zeta (3) \zeta (5) + \frac {1953}{8} \pi ^2 \ln 2 \zeta (3) \zeta (5) - \nonumber \\
&-& \frac {341}{8} \pi ^4 \ln 2 (2\ln 2 +\gamma _E)\zeta (5)-\frac {93}{4} \pi ^4 (\ln 2)^2\zeta (5) -
 \frac {385}{48} \pi ^4 (2\ln 2 +\gamma _E)\zeta (3)^2- \frac {413}{12} \pi ^4 \ln 2\zeta (3)^2 + \nonumber \\
 &+&  \frac {343}{8} \pi ^2 \zeta (3)^3 + \frac {767}{720} \pi ^6 \ln 2 (2\ln 2 +\gamma _E)\zeta (3)+
  \frac {35}{12}\pi ^6 (\ln 2)^2 \zeta (3) \, .
\end{eqnarray}

\section{Systematics of the subleading term}
\setcounter{equation}{0}

We now want to put the linear integral equation (\ref {Skeq}) in a
form which is more suitable for analysis of both the weak and the
strong coupling limit. In brief, instead of working with one
linear integral equation (\ref {Skeq}), we will be concerned with
(two) linear infinite (i.e. containing infinite equations)
systems.

As far as the dependence on $\ln s$ is concerned, equation (\ref
{Skeq}) and its solution split in a part proportional to $\ln s$
and a part proportional to $(\ln s)^0$.

We concentrate on the latter, which we call $S^{(0)}(k)$, since
the former has been extensively studied \cite {FRS,FGR3}.

We now restrict to the domain $k \geq 0$ and expand $S^{(0)}(k)$ in series of
Bessel functions
\begin{equation}
S^{(0)}(k)=\sum _{r=1}^{\infty} S^{(0)}_r(g) \frac {J_r({\sqrt
{2}}gk)}{k} \, . \label {sorg}
\end{equation}
As a consequence of (\ref {Skeq}), the coefficients $S^{(0)}_r(g)$
satisfy the following system of equations,
\begin{eqnarray}
S^{(0)}_{2p-1}(g)&=&2 {\sqrt {2}}g \gamma _E \delta _{p,1} +
4(2p-1)\int _{0}^{\infty}
\frac {dh}{h} \frac {\tilde J_{2p-1}({\sqrt {2}}gh)}{e^h-1}+A_{2p-1}^{(0)}(g)
-2(2p-1)
\sum _{m=1}^{\infty} Z_{2p-1,m}(g) S^{(0)}_m (g) \nonumber \\
\label {S0system} \\
S^{(0)}_{2p}(g)&=&4+ 8p \int _{0}^{\infty}
\frac {dh}{h} \frac {J_{2p}({\sqrt {2}}gh)}{e^h-1}+A_{2p}^{(0)}(g)
+4p\sum _{m=1}^{\infty} Z_{2p,2m-1}(g) S^{(0)}_{2m-1}(g)
-4p\sum _{m=1}^{\infty} Z_{2p,2m}(g) S^{(0)}_{2m}(g) \, , \nonumber
\end{eqnarray}
where, as usual, we introduced the notation
\begin{equation}
Z_{n,m}(g)= \int _{0}^{\infty}
\frac {dh}{h} \frac {J_{n}({\sqrt {2}}gh)J_{m}({\sqrt {2}}gh)}{e^h-1}
\, .
\end{equation}
and where we introduced the function $\tilde J_{2p-1}(x)$ which coincides with
the Bessel function $J_{2p-1}(x)$ for $p\geq 2$ and with $J_1(x)-\frac {x}{2}$
when $p=1$.
In (\ref {S0system}) the terms $A_r^{(0)}(g)$ is the term proportional to $(\ln s)^{0}$ of the quantity:
\begin{equation}
A_r(g)=r \int _{0}^{+\infty}\frac {dh}{2\pi h} \,  \frac
{J_{r}({\sqrt {2}}gh)}{\sinh \frac {h}{2}} \,  \int
_{-\infty}^{+\infty} \frac {dp}{2\pi} 2 \left [ \frac {\sin (h-p)c }{h-p}-
\frac {\sin pc}{p} \right ] \hat \sigma (p)   \, .
\label {forcrn}
\end{equation}
Eventually, anomalous dimension at order $(\ln s)^0$, $\gamma
^{(0)}$, is extracted from $S_1^{(0)}(g)$ by means of the formula
\begin{equation}
\gamma ^{(0)}={\sqrt {2}}g S_1^{(0)}(g) \, .
\end{equation}
To be complete, we write down also the equations satisfied by the
part of $S(k)$, which is linear in $\ln s$. Let us call such part
$S^{(-1)}(k)$. Expanding in series of Bessel functions
\begin{equation}
S^{(-1)}(k)=\sum _{p=1}^{\infty} S^{(-1)}_p(g)
\frac {J_p({\sqrt {2}}gk)}{k} \, ,
\end{equation}
we find the following system of equations for the coefficients
\begin{eqnarray}
S^{(-1)}_{2p-1}(g)&=&2 {\sqrt {2}}g \delta _{p,1} - 2(2p-1)\, j \,
\int _{0}^{\infty} \frac {dh}{h} \frac {J_{2p-1}({\sqrt
{2}}gh)}{e^{\frac {h}{2}}+1}+A_{2p-1}^{(-1)}(g) - \nonumber \\
&-& 2(2p-1)
\sum _{m=1}^{\infty} Z_{2p-1,m}(g) S^{(-1)}_m (g) \nonumber \\
\label {Ssystem} \\
S^{(-1)}_{2p}(g)&=&2j - 4p \, j \, \int _{0}^{\infty} \frac
{dh}{h} \frac {J_{2p}({\sqrt {2}}gh)}{e^{\frac {h}{2}}+1}+
 A_{2p}^{(-1)}(g) + \nonumber \\
 &+& 4p\sum _{m=1}^{\infty} Z_{2p,2m-1}(g)
S^{(-1)}_{2m-1}(g) -4p\sum _{m=1}^{\infty} Z_{2p,2m}(g)
S^{(-1)}_{2m}(g) \, , \nonumber
\end{eqnarray}
where $A_r^{(-1)}(g)$ is the term proportional to $\ln s$ of the quantity
(\ref {forcrn}).

This system has been used in the series of papers \cite
{FGR1,FGR2,FGR3} in order to compute the (strong coupling limit
of) the generalised scaling function $f(g,j)$. In the next
subsections we will adapt the steps of \cite {FGR1,FGR2,FGR3} to
system (\ref {S0system}) in order to study the subleading
function $f^{(0)}(g,j)$ as series in $j$, up to the order $j^5$.

\subsection{\bf Slicing in powers of $j$}

Since we are in the limit (\ref {jlimit}), the function
$S^{(0)}(k)$ admits an expansion in powers of $j$,
\begin{equation}
S^{(0)}(k)=\sum _{n=0}^{\infty} S^{(0,n)}(k)j^n \, . \label{s0exp}
\end{equation}
Consequently, this way of expanding extends to the coefficients
$S^{(0)}_r(g)$ (\ref {sorg}) as well:
\begin{equation}
S^{(0)}_r(g)=\sum _{n=0}^{\infty} S^{(0,n)}_r (g) j^n \, . \label {sorg2}
\end{equation}

In expansion (\ref {sorg2}) the coefficients $S^{(0,0)}_r(g)$ -
which give the part of the function $S^{(0)}(k)$ independent of
$j$ - satisfy the system of equations
\begin{eqnarray}
S^{(0,0)}_{2p-1}(g)&=&2 {\sqrt {2}}g \gamma _E \delta _{p,1} +
4(2p-1)\int _{0}^{\infty}
\frac {dh}{h} \frac {\tilde J_{2p-1}({\sqrt {2}}gh)}{e^h-1}
-2(2p-1)
\sum _{m=1}^{\infty} Z_{2p-1,m}(g) S^{(0,0)}_m (g) \nonumber \\
\label {S00system} \\
S^{(0,0)}_{2p}(g)&=&4+ 8p \int _{0}^{\infty} \frac {dh}{h} \frac
{J_{2p}({\sqrt {2}}gh)}{e^h-1} +4p\sum _{m=1}^{\infty}
Z_{2p,2m-1}(g) S^{(0,0)}_{2m-1}(g) -4p\sum _{m=1}^{\infty}
Z_{2p,2m}(g) S^{(0,0)}_{2m}(g) \, . \nonumber
\end{eqnarray}
This system of equations is studied in the contemporaneous paper
\cite {FGR4} - the coefficients $S^{(0,0)}_r(g)$ being there
denoted as $S^{extra}_r(g)$. Therefore, we pass to study the
function $S^{(0,n)}_r(g)$, when $n \geq 1$. For simplicity's sake
we limit to the cases $n=3,4,5$ (for $n=1,2$, $S^{(0,n)}_r(g)=0$).
The relevant forcing terms are obtained after developing (\ref
{forcrn}) for small $c$ up to $c^5$:
\begin{equation}
A_r^{(0,n)}(g)=r \int _{0}^{+\infty}\frac {dh}{2\pi h} \frac
{J_{r}({\sqrt {2}}gh)}{\sinh \frac {h}{2}} \,  \left [ -\frac {1}{3} h^2
c^3 \sigma (0) +\frac {1}{60}h^4c^5 \sigma (0)-\frac {1}{10}h^2 c^5 \sigma _2 (0) \right ]_{(\ln s)^0,j^n}  \, , \quad 3 \leq n \leq 5 \, .
\label {forcrn3}
\end{equation}
Now, from the expansions in powers of $\ln s$:
\begin{equation}
\sigma (0)=\ln s \ \sigma ^{(-1)}(0)+ \sigma ^{(0)}(0) +
O\left (\frac {1}{\ln s}\right ) \, , \quad
c=c^{(0)}+(\ln s )^{-1} \ c^{(1)}+ O\left (\frac {1}{(\ln s)^2}\right )\, ,
\end{equation}
we get ($3 \leq n \leq 5$):
\begin{eqnarray}
A_r^{(0,n)}(g)&=&r \int _{0}^{+\infty}\frac {dh}{2\pi h} \frac
{J_{r}({\sqrt {2}}gh)}{\sinh \frac {h}{2}} \,  \Bigl [ -\frac {1}{3} h^2
\left ( {c^{(0)}}^3 \sigma ^{(0)}(0) + 3 {c^{(0)}}^2  c^{(1)} \sigma ^{(-1)}(0) \right ) + \label {forcrn4} \\
&+& \frac {1}{60}h^4 \left ( {c^{(0)}}^5 \sigma ^{(0)}(0) + 5 {c^{(0)}}^4  c^{(1)} \sigma ^{(-1)}(0) \right )-\frac {1}{10}h^2\left ( {c^{(0)}}^5 \sigma ^{(0)}_2(0) + 5 {c^{(0)}}^4  c^{(1)} \sigma ^{(-1)}_2 (0) \right )  \Bigr ]_{j^n}  \, .\nonumber
\end{eqnarray}
Keeping in mind the expansions in powers of $j$:
\begin{equation}
c^{(k)}=\sum _{n=1}^{\infty} c^{(k,n)}j^n \, , \quad \sigma ^{(k-1)}(0)=
\sum _{n=0}^{\infty}\sigma ^{(k-1,n)}(0)j^n \, , \quad k=0,1 \, ,
\end{equation}
we can now specialise (\ref {forcrn4}) to the cases $n=3$,
\begin{equation}
A_r^{(0,3)}(g)=r \int _{0}^{+\infty}\frac {dh}{2\pi h} \frac
{J_{r}({\sqrt {2}}gh)}{\sinh \frac {h}{2}} \left [ -\frac {1}{3} h^2
\left ( {c^{(0,1)}}^3 \sigma ^{(0,0)}(0) + 3 {c^{(0,1)}}^2  c^{(1,1)} \sigma ^{(-1,0)}(0) \right )\right ]
\label {ar3} \, ,
\end{equation}
$n=4$,
\begin{eqnarray}
A_r^{(0,4)}(g)&=&r \int _{0}^{+\infty}\frac {dh}{2\pi h} \frac
{J_{r}({\sqrt {2}}gh)}{\sinh \frac {h}{2}} \Bigl [ -\frac {1}{3} h^2
\Bigl ( 3 {c^{(0,1)}}^2 c^{(0,2)}\sigma ^{(0,0)}(0) + 3 {c^{(0,1)}}^2  c^{(1,1)} \sigma ^{(-1,1)}(0) + \nonumber \\
&+&  3 {c^{(0,1)}}^2  c^{(1,2)} \sigma ^{(-1,0)}(0)
+6{c^{(0,1)}} c^{(0,2)} c^{(1,1)} \sigma ^{(-1,0)}(0) \Bigr )\Bigr ]
\label {ar4}
\end{eqnarray}
and $n=5$:
\begin{eqnarray}
A_r^{(0,5)}(g)&=&r \int _{0}^{+\infty}\frac {dh}{2\pi h} \frac
{J_{r}({\sqrt {2}}gh)}{\sinh \frac {h}{2}} \Bigl [ -\frac {1}{3} h^2
\Bigl ( 3 {c^{(0,1)}}^2 c^{(0,3)}\sigma ^{(0,0)}(0) + 3 {c^{(0,2)}}^2  c^{(0,1)} \sigma ^{(0,0)}(0) + \nonumber \\
&+&  3 {c^{(0,1)}}^2  c^{(1,3)} \sigma ^{(-1,0)}(0)
+6{c^{(0,1)}} c^{(0,2)} c^{(1,2)} \sigma ^{(-1,0)}(0)
+6{c^{(0,1)}} c^{(0,3)} c^{(1,1)} \sigma ^{(-1,0)}(0) + \nonumber \\
&+&  3 {c^{(0,2)}}^2  c^{(1,1)} \sigma ^{(-1,0)}(0)+
3 {c^{(0,1)}}^2  c^{(1,2)} \sigma ^{(-1,1)}(0)+6{c^{(0,1)}} c^{(0,2)} c^{(1,1)} \sigma ^{(-1,1)}(0) \Bigr ) + \nonumber \\
&+& \frac {1}{60}h^4 \left ( {c^{(0,1)}}^5 \sigma ^{(0,0)}(0) + 5 {c^{(0,1)}}^4  c^{(1,1)} \sigma ^{(-1,0)}(0) \right )- \nonumber \\
&-& \frac {1}{10}h^2\left ( {c^{(0,1)}}^5
\sigma ^{(0,0)}_2(0) + 5 {c^{(0,1)}}^4  c^{(1,1)} \sigma ^{(-1,0)}_2 (0)
\right ) \Bigr ] \, .
\label {ar5}
\end{eqnarray}
On the other hand, the $c$ can be expressed in terms of the $\sigma (0)$ by
means of the condition (\ref {scond}). At the relevant order in $c$ such condition reads
\begin{equation}
2\sigma (0) c +\frac {1}{3} \sigma _2 (0) c^3 =-2 \pi j \ln s \, .
\label {sigmac}
\end{equation}
The order $\ln s ^0$ of such equation gives
\begin{equation}
2  \sigma ^{(-1)}(0)c^{(1)}+2  \sigma ^{(0)}(0)c^{(0)}+
\sigma ^{(-1)}_2(0){c^{(0)}}^2c^{(1)}+ \frac {1}{3} \sigma ^{(0)}_2(0)
{c^{(0)}}^3=0
\end{equation}
Specialising such equation at order $j$, we get the condition \footnote {We
remember the formul{\ae} (relations (5.18-5.20) of \cite {FGR3}):
\begin{equation}
c^{(0,1)}=-\frac {\pi} {\sigma ^{(-1,0)}(0)} \, , \quad
c^{(0,2)}=\pi \frac {\sigma ^{(-1,1)}(0)} {[\sigma ^{(-1,0)}(0)]^2} \, , \quad
c^{(0,3)}=\frac {\pi ^3}{6} \frac {\sigma ^{(-1,0)}_2(0)} {[\sigma ^{(-1,0)}(0)]^4}-\pi \frac {[\sigma ^{(-1,1)}(0)]^2} {[\sigma ^{(-1,0)}(0)]^3}  \, .
\end{equation}}
\begin{equation}
c^{(1,1)}=-\frac { \sigma ^{(0,0)}(0)}{\sigma ^{(-1,0)}(0)}c^{(0,1)} =
\pi \frac { \sigma ^{(0,0)}(0)}{[\sigma ^{(-1,0)}(0)]^2} \, .
\end{equation}
At order $j^2$ we have
\begin{equation}
c^{(1,2)}=-\frac { \sigma ^{(-1,1)}(0)}{\sigma ^{(-1,0)}(0)}c^{(1,1)}
-\frac { \sigma ^{(0,0)}(0)}{\sigma ^{(-1,0)}(0)}c^{(0,2)}
= -2\pi \frac { \sigma ^{(0,0)}(0)\sigma ^{(-1,1)}(0)}
{[\sigma ^{(-1,0)}(0)]^3} \, .
\end{equation}
Going at order $j^3$ we get
\begin{eqnarray}
c^{(1,3)}&=&-\frac { \sigma ^{(-1,1)}(0)}{\sigma ^{(-1,0)}(0)}c^{(1,2)}
-\frac { \sigma ^{(0,0)}(0)}{\sigma ^{(-1,0)}(0)}c^{(0,3)}-\frac {1}{2}
\frac {\sigma ^{(-1,0)}_2(0)} {\sigma ^{(-1,0)}(0)}{c^{(0,1)}}^2 c^{(1,1)}
-\frac {1}{6}
\frac {\sigma ^{(0,0)}_2(0)} {\sigma ^{(-1,0)}(0)}{c^{(0,1)}}^3 = \nonumber \\
&=& 3 \pi  \frac { \sigma ^{(0,0)}(0)[\sigma ^{(-1,1)}(0)]^2}
{[\sigma ^{(-1,0)}(0)]^4}-\frac {2}{3} \pi ^3
\frac { \sigma ^{(0,0)}(0)\sigma ^{(-1,0)}_2(0)}
{[\sigma ^{(-1,0)}(0)]^5}+\frac {\pi ^3}{6} \frac {\sigma ^{(0,0)}_2(0)} {[\sigma ^{(-1,0)}(0)]^4}\, .
\end{eqnarray}
In such a way, we obtain
\begin{equation}
A_r^{(0,3)}(g)=r \int _{0}^{+\infty}\frac {dh}{2\pi h} \frac
{J_{r}({\sqrt {2}}gh)}{\sinh \frac {h}{2}} \,  \left (-\frac {2}{3} \pi ^3 h^2 \right )
\frac { \sigma ^{(0,0)}(0)} {[\sigma ^{(-1,0)}(0)]^3 }  \, .
\label {Ar03}
\end{equation}
and, also,
\begin{equation}
A_r^{(0,4)}(g)=r \int _{0}^{+\infty}\frac {dh}{2 \pi h} \frac
{J_{r}({\sqrt {2}}gh)}{\sinh \frac {h}{2}} \,  2 \pi ^3  h^2
\frac { \sigma ^{(0,0)}(0)\sigma ^{(-1,1)}(0)} {[\sigma ^{(-1,0)}(0)]^4 }  \, .
\label {Ar04}
\end{equation}
For what concerns the term proportional to $j^5$, we get, after some
calculation:
\begin{eqnarray}
A_r^{(0,5)}(g)&=&r \int _{0}^{+\infty}\frac {dh}{2\pi h} \frac
{J_{r}({\sqrt {2}}gh)}{\sinh \frac {h}{2}} \Bigl [ -\frac {1}{3} h^2
\Bigl ( \frac {\pi ^5}{5}
\frac {\sigma _2^{(0,0)}(0)}{[\sigma ^{(-1,0)}(0)]^5} + \nonumber \\
&+&  12 \pi ^3  \frac {\sigma ^{(0,0)}(0)[\sigma ^{(-1,1)}(0)]^2}
{[\sigma ^{(-1,0)}(0)]^5} -\pi ^5  \frac {\sigma ^{(0,0)}(0)
\sigma ^{(-1,0)}_2(0)}
{[\sigma ^{(-1,0)}(0)]^6} \Bigr ) +
\frac {1}{15}h^4 \pi ^5   \frac {\sigma ^{(0,0)}(0)}{[\sigma ^{(-1,0)}(0)]^5}
\Bigr ] \label {Ar05}
\end{eqnarray}
Now, in analogy to what done in \cite {FGR3}, we introduce the
'reduced' coefficients $\tilde S_r^{(k)}(g)$ which satisfy the
system (4.23) of \cite {FGR3}, i.e. the 'usual' system with the
BES kernel and with forcing terms
\begin{equation}
{\mathbb I}_r^{(k)}(g)=r \int _{0}^{+\infty}\frac {dh}{2 \pi }
h^{2k-1} \frac {J_{r}({\sqrt {2}}gh)}{\sinh \frac {h}{2}}  \, .
\label {Irk}
\end{equation}
Using notations of \cite {FGR3}, we can write the various
$f^{(0)}_n$, $n=3,4,5$, in terms of $\tilde S_1^{(k)}(g)$ and of
the density and its derivatives in zero
\begin{eqnarray}
f_3^{(0)}(g)&=&-\frac {2}{3} \pi ^3  \frac { \sigma ^{(0,0)}(0)}
{[\sigma ^{(-1,0)}(0)]^3 } {\sqrt {2}}g \tilde S_1^{(1)}(g) \, , \label {f03} \\
f_4^{(0)}(g)&=&  2 \pi ^3   \frac { \sigma ^{(0,0)}(0)\sigma
^{(-1,1)}(0)} {[\sigma ^{(-1,0)}(0)]^4 } {\sqrt {2}}g \tilde
S_1^{(1)}(g) \, , \label {f04} \\
f_5^{(0)}(g)&=& \Bigl [ -\frac {1}{3} \Bigl ( \frac {\pi ^5}{5}
\frac {\sigma _2^{(0,0)}(0)}{[\sigma ^{(-1,0)}(0)]^5} + 12 \pi ^3
\frac {\sigma ^{(0,0)}(0)[\sigma ^{(-1,1)}(0)]^2} {[\sigma
^{(-1,0)}(0)]^5} -\pi ^5  \frac {\sigma ^{(0,0)}(0) \sigma
^{(-1,0)}_2(0)} {[\sigma ^{(-1,0)}(0)]^6} \Bigr ) {\sqrt {2}}g
\tilde S_1^{(1)}(g) + \nonumber \\
&+& \frac {\pi ^5}{15}    \frac {\sigma ^{(0,0)}(0)}{[\sigma
^{(-1,0)}(0)]^5} {\sqrt {2}}g \tilde S_1^{(2)}(g)\Bigr ] \, .
\label {f05}
\end{eqnarray}
For comparison, corresponding formul{\ae} for the coefficients
$f_n(g)$ of the function $f(g,j)$ are, (5.30-32) of \cite {FGR3},
\begin{eqnarray}
f_3(g)&=&\frac {\pi ^3}{3}   \frac {1}
{[\sigma ^{(-1,0)}(0)]^2 } {\sqrt {2}}g \tilde S_1^{(1)}(g) \, , \label {f3} \\
f_4(g)&=& - \frac {2}{3} \pi ^3   \frac { \sigma ^{(-1,1)}(0)}
{[\sigma ^{(-1,0)}(0)]^3 } {\sqrt {2}}g \tilde
S_1^{(1)}(g) \, , \label {f4} \\
f_5 (g)&=&\Bigl [ \pi ^3  \frac {[\sigma ^{(-1,1)}(0)]^2}{[\sigma
^{(-1,0)}(0)]^4} - \frac {\pi ^5}{15} \frac {\sigma
^{(-1,0)}_2(0)} {[\sigma ^{(-1,0)}(0)]^5}  \Bigr ] {\sqrt {2}}g
\tilde S_1^{(1)}(g) - \frac {\pi ^5}{60}     \frac {1}{[\sigma
^{(-1,0)}(0)]^4} {\sqrt {2}}g \tilde S_1^{(2)}(g) \, . \label {f5}
\end{eqnarray}
Expressions (\ref {f03}, \ref {f04}, \ref {f05}) interpolate from
weak to strong coupling and for the moment lack of an explicit
form as functions of $g$. Their weak coupling expansion in powers
of $g^2$ were given, up to $g^6$, in Section 3. In the next
subsection, we will study and explicitly find their strong
coupling limit.

\subsection{Strong coupling}

It is interesting to consider the strong coupling limit of the
equations obtained in the last subsection. For that purpose, we
need to know the values of the density and its derivatives in zero
and the quantities $\tilde S_1^{(k)}(g)$ as well. Results
concerning $\sigma ^{(-1,n)}$, i.e. the part of the density
proportional to $\ln s$, and $\tilde S_1^{(k)}(g)$ are reported in
\cite {FGR3}. For what concerns the contribution to the anomalous
dimension proportional to $\ln s$, i.e. $f(g,j)$, we remember
\cite {AM,BK} that in the limit (\ref {jlimit}) and when $j \ll g$
it coincides with the energy density of the $O(6)$ sigma model.
The nonperturbative (infrared) regime of the sigma model, $j \ll
m(g)$, where
\begin{equation}
m(g)= k g^{\frac {1}{4}}e^{-\frac {\pi g }{\sqrt {2}}} + \ldots \, , \quad
k= \frac {2^{\frac {5}{8}}\pi}{\Gamma \left (\frac {5}{4}\right )}
\, ,
\end{equation}
where the dots stand for subleading corrections, 
makes contact with the double limit $j \ll 1$, $ g \rightarrow
\infty$, considered in this subsection (and also in \cite {FGR3}).
In such regime the scale of the mass is given by $m(g)$ and
expansion of $f(g,j)$ for small $j$ and large $g$ was successfully
checked \cite {FGR2,FGR3} against analogous non perturbative
expansions \cite {BF} in the sigma model.

It remains an open question if, as well as $f(g,j)$, also
$f^{(0)}(g,j)$ shows connections with quantities of the $O(6)$
sigma model. This is a further motivation which leads to study the
strong coupling limit of $f^{(0)}(g,j)$, which however is an
important problem in itself, since its results can be checked
against string theory data. We concentrate on $f^{(0)}_n(g)$, for
$n=3,4,5$: as follows from (\ref {f03}, \ref {f04}, \ref {f05}),
the calculation of their strong coupling limit can be finalised if
we know the large $g$ behaviour of the density and its second
derivative in zero, $\sigma ^{(0,0)}(0)$ and $\sigma
^{(0,0)}_2(0)$. From results of \cite {FGR4}, we know that at
large $g$ and at the leading order
\begin{equation}
S_r^{(0,0)}(g)=-\ln g \, S_r^{(-1,0)}(g)+\ldots \, ,
\end{equation}
where $S^{(0,0)}_p(g)$ is a solution of (\ref {S00system}) and
$S^{(-1,0)}_p(g)$ of (\ref {Ssystem}) with $j=0$ (i.e. the BES
system). From this relation we can deduce that, at the leading
order, the density in zero reads as
\begin{equation}
\sigma _H^{(0,0)}(0)=-\ln g \, \sigma _H^{(-1,0)}(0)+\ldots = - \ln g [ 4-\pi m(g)]+ \ldots \Rightarrow \sigma ^{(0,0)}(0)=-4 \ln g + \ldots
\end{equation}
Analogously, for what concerns the second derivative, we get
\begin{equation}
\sigma _2^{(0,0)}(0)=56 \zeta (3)+ \ldots
\end{equation}
since the higher loops contributions, $\frac {\pi ^3}{4}\ln g \,
m(g)$, are exponentially depressed.

Using the results contained in \cite {FGR3}, we get the strong
coupling relations (which are valid only at the leading order),
\begin{eqnarray}
f_3^{(0)}(g)&=&-\frac {8 \ln g}{\pi m(g)} f_3(g)+...=- \frac {\pi \ln g}{3 m(g)^2}+...\\
f_4^{(0)}(g)&=&-\frac {12 \ln g}{\pi m(g)} f_4(g)+...=\frac {\ln g}{m(g)^3}\left (\ln 2 + \frac {\pi}{2}\right )+...\\
f_5^{(0)}(g)&=& \Bigl [ -\frac {16 \ln g}{\pi ^2 m(g)^5} \Bigl
(\ln 2 +\frac {\pi }{2}\Bigr )^2 + \frac {\pi ^2 \ln g }{3 m(g)^5}
\Bigr ] {\sqrt {2}}g
\tilde S_1^{(1)}(g) + \nonumber \\
&+& \frac {16 \ln g}{60 m(g)^5}   {\sqrt {2}}g \tilde
S_1^{(2)}(g)+...= -\frac {16 \ln g}{\pi  m(g)}f_5(g)+\frac {\pi ^2
\ln g}{15 m(g)^5}{\sqrt {2}}g \tilde S_1^{(1)}(g)+...= \nonumber
\\
&=& -\frac {2 \ln g}{\pi  m(g)^4}\left (\ln 2 + \frac
{\pi}{2}\right )^2+ \frac {\pi ^3 \ln g}{30  m(g)^4}+...
\end{eqnarray}
We see that there is a simple proportionality between
$f_n^{(0)}(g)$ and $(n-1) f_n(g)$ for $n=3,4$, which, however, is
lost when $n=5$. This could suggest a relation between
$f^{(0)}(g,j)$ and $f(g,j)$, but at the moment our data are not
conclusive about its form.

\section{Conclusions}

A systematic procedure for the computation of the subleading
correction $f^{(0)}(g,j)$ in the high spin limit (\ref {jlimit})
to the anomalous dimensions in the $sl(2)$ sector of ${\cal N}=4$
SYM is developed. The method is analogous to the one used in \cite
{FGR3} to study the leading term $f(g,j)$ and is based on a linear
integral equation describing the behaviour of the density and the
observables at high spin. We first found one loop (subleading)
contributions to the anomalous dimension (Section 2), finding
agreement with results obtained by another method, which is an
improvement of the techniques of \cite {BGK} (Appendix B). Then,
we performed weak coupling expansions (two and three loops,
Section 3). Finally, we wrote the linear integral equation for the
density as linear infinite systems (Section 4). This was
particularly convenient, for it simplifies the study of the strong
coupling limit of $f^{(0)}(g,j)$. In this respect, we explicitly
found (Subsection 4.2) the large $g$ limit of $f_n^{(0)}(g)$, for
$n=3,4,5$, comparing their expression with $f_n(g)$. Although the string action may reduce to the $O(6)$ non-linear sigma model, the question of its appearance in the sub-leading scaling function still stays as an open question.

\vspace {1.5cm}

{\bf Acknowledgements} DF  ought to particularly thank  A. Tseytlin for useful discussions and suggestions. Moreover, we thank D.
Bombardelli, P. Grinza for discussions and suggestions. We
acknowledge the INFN grant {\it Iniziative specifiche FI11} and
{\it PI14}, the italian University PRIN 2007JHLPEZ "Fisica
Statistica dei Sistemi Fortemente Correlati all'Equilibrio e Fuori
Equilibrio: Risultati Esatti e Metodi di Teoria dei Campi" for
travel financial support. DF acknowledges the Galileo Galilei
Institute for Theoretical Physics as well as  M.R.  the
INFN/University of Bologna for hospitality.

\appendix

\section{Estimate of the holes contribution}
\setcounter{equation}{0}

We now give an estimate of the sum over the internal holes $u_h$
of a generic function $O(u_h)$. Results we present here are valid
both at one and all loops cases. We start from the exact
hole expression \cite{FMQR}, depending on the so-called counting function $Z(u)$
\footnote {The counting function is connected to
$\sigma (u)$ by the relation $Z^{\prime}(u)=\sigma (u)$.},
\begin{eqnarray}
\sum _{h=1}^{L-2} O(u_h)&=&-\int _{-c}^{c} \frac {du}{2\pi} O(u)
Z^{\prime}(u)+ {\mbox {Im}} \int _{-c}^{c}\frac {dv}{\pi}
O(v-i\epsilon)\frac {d}{dv} \ln
[1+\delta \, e^{iZ(v-i\epsilon)}] + \label {int} \\
&+& {\mbox {Im}} \int _{0}^{-\epsilon }\frac {dy}{\pi}
O(-c+iy)\frac {d}{dy} \ln [1+\delta \, e^{iZ(-c+iy)}]+ {\mbox
{Im}} \int _{-\epsilon }^{0}\frac {dy}{\pi} O(c+iy)\frac {d}{dy}
\ln [1+\delta \, e^{iZ(c+iy)}] \, . \nonumber
\end{eqnarray}
The right hand side of (\ref {int}) does not depend on $\epsilon $
as far as no poles of the integrands lie in the region ${\mbox
{Im}}v < \epsilon$, $|{\mbox {Re}}v| <c$. The constant $\delta $
is equal to $\pm 1$, depending on the parity of $L$. Keeping
$\epsilon$ small, but finite, we suppose that
\begin{equation}
\left | e^{iZ(z)}\right | \ll 1 \, , \label {appr}
\end{equation}
when $z$ belongs to the integration contour of (\ref {int}). In
the large $s$ limit this is justified, since $Z^{\prime}(v)$ is
proportional to $-\ln s$. Using such approximation, we can replace
all the $\ln [1+e^{iZ(z)}]$ with $e^{iZ(z)}$. Then, since the
expression we get is still independent of $\epsilon$, we find
convenient to evaluate it when $\epsilon =0$ \footnote {We can put
directly $\epsilon =0$ since, after the approximation (\ref
{appr}), the integrand is regular on the real axis.}. We obtain
\begin{equation}
\sum _{h=1}^{L-2} O(u_h)=-\int _{-c}^{c} \frac {du}{2\pi} O(u)
Z^{\prime}(u)+ \delta \, {\mbox {Im}} \int _{-c}^{c}\frac
{dv}{\pi} O(v)\frac {d}{dv} e^{iZ(v)} \, . \label{sumh}
\end{equation}
The nonlinear term in this expression
\begin{equation}
NL=\delta \, {\mbox {Im}} \int _{-c}^{c}\frac {dv}{\pi} O(v)\frac
{d}{dv} e^{iZ(v)} \, ,
\end{equation}
can be estimated after the change of variable $x=Z(v)$ (we remind
that $Z^{\prime}(v)<0$) and after using the formula
\begin{equation}
\int dx f(x) e^{ax}=\frac {e^{ax}}{a} \sum _{k=0}^{\infty} (-1)^k
\frac {f^{(k)}(x)}{a^k} \, ,
\end{equation}
where $f^{(k)}(x)$ denotes the $k$-th derivative of $f(x)$. We get
\begin{equation}
NL= \delta \, {\mbox {Im}} \left [ \frac {e^{ix}}{\pi} \sum
_{k=0}^{\infty} i^k \frac {\partial} {\partial  x^k} O(Z^{-1}(x))
\right ] _{Z(-c)}^{Z(c)} \, . \label {NLseries}
\end{equation}
The first terms of such series are
\begin{equation}
NL=\delta \, \frac {2}{\pi} O(c) \sin Z(c) + \delta \, \frac
{2}{\pi} \frac {O^{\prime}(c)} {Z^{\prime}(c)} \cos Z(c) +\ldots
\label {NL}
\end{equation}
where the dots represent terms containing higher powers of
$Z^{\prime}(c)$ in the denominator. Now, if $c$ is chosen such
that $Z(c)=-\pi (L-2)+O(1/\ln s)$ - which is exactly condition
(\ref {s0cond}) for the one loop and (\ref {scond}) for the all
loops case, respectively - the first term in (\ref {NL}) is
$O(1/\ln s)$. The second term is $O(1/\ln s)$ as well, since
$Z^{\prime}(c)$ - for generic $g$ - is proportional to $\ln s$.
After carefully evaluating all the terms in the series, one can
show by similar reasonings that the $k$-th term is $O(1/(\ln
s)^k)$: this ensures the convergence of the series in (\ref
{NLseries}), which gives the nonlinear contribution to (\ref
{sumh}). Evidently, such nonlinear contribution is $O(1/\ln s)$.

\section{Contact with \cite {BGK}}
\setcounter{equation}{0}

We now show that our one loop results of Section 1 can be obtained
starting from the equations of \cite {BGK}. In this paper, authors
are interested to the contribution to the energy proportional to
$\ln s$ (i.e. the part $\ln s f(j)$, where $f(j)$ is the one loop
contribution to the cusp anomalous dimension). However, their
equations can be used to compute also the subleading part,
proportional to $\ln s ^0$, both at the level of equation for the
density and at the level of explicit computations of energy
eigenvalues.

Let us start from the equations describing the distribution of
roots and holes. In paper \cite {BGK}, authors start from the
equation
\begin{equation}
2 \delta _n \ln s - \frac {iL}{2} \ln \frac {\Gamma \left (\frac
{1}{2}-i\delta _n \right )} {\Gamma \left (\frac {1}{2}+i\delta _n
\right )} -\frac {i}{2} \sum _{j=2}^{L-1} \ln  \frac {\Gamma \left
(1+i\delta _n -i\delta _j \right )} {\Gamma \left (1-i\delta _n
+i\delta _j \right )}=\frac {\pi}{2} k_n \, , \label {A1}
\end{equation}
which describe the distribution of internal holes, denoted as
$\delta _n$, $2\leq n\leq L-1$. This equation suggests the
definition of the counting function
\begin{equation}
Z_0(u)=-4\ln s \, u + iL \ln \frac {\Gamma \left (\frac {1}{2}-iu
\right )}
 {\Gamma \left (\frac {1}{2}+iu \right )}+ i \sum _{j=2}^{L-1} \ln
\frac {\Gamma (1+iu - i \delta _j)}{\Gamma (1-iu - i \delta _j)}
\, . \label {Zholes}
\end{equation}
Indeed we have that $Z_0(\delta _n)=\pi (2n-L-1)\, , \quad e^{i
Z_0(\delta _n)}=(-1)^{L+1}$.

In terms of the derivative $\sigma _0(u)= \frac {d}{du} Z_0(u)$,
we have the equation
\begin{equation}
\sigma _0(u)=-4\ln s +L\left [ \psi \left (\frac {1}{2}-iu \right
)+
 \psi \left (\frac {1}{2}+iu \right )\right ] - \sum _{j=2}^{L-1}[
\psi (1+iu -i \delta _j)+  \psi (1-iu -i \delta _j)] \, ,
\end{equation}
which after expressing the sum on the holes in terms of the
density of holes $-\frac {1}{2\pi} \sigma _0(u)$ turns out into
\begin{equation}
\sigma _0(u)=-4\ln s +L\left [ \psi \left (\frac {1}{2}-iu \right
)+
 \psi \left (\frac {1}{2}+iu \right )\right ]
+\int _{-c_0}^{c_0}\frac {dv}{2\pi}\sigma _0 (v) [ \psi (1+iu -i
v)+  \psi (1-iu -i v)] \, . \label {sigmaholes}
\end{equation}
Such equation describes the distribution of holes and,
consequently, it is valid only for $-c_0 \leq u \leq c_0$. And,
indeed, one can verify that our starting equation (\ref {S0u})
reduces exactly to (\ref {sigmaholes}) when $-c_0 \leq u \leq
c_0$.

After having shown that our equation for the density of roots and
holes agrees with the corresponding relation that can be deduced
from results of \cite {BGK}, we pass to compare results for the
eigenvalues of the energy: we verify that expression (\ref {E0})
can be obtained starting from formul{\ae} of \cite {BGK}.

Referring still to equation (\ref {A1}), describing the
distribution of internal holes, we develop such equation up to the
order $\delta _n^3$. We obtain the relation
\begin{equation}
2 \delta _n \ln s-L \psi \left (\frac {1}{2}\right ) \delta _n +\frac {L}{6} \psi ^{(2)}\left (\frac {1}{2}\right )
\delta _n ^3 +(L-2) \psi (1) \delta _n - \frac {L-2}{6}\psi ^{(2)}(1) \delta _n ^3 - \frac {1}{2}\psi ^{(2)}(1) \delta _n \sum _{j=2}^{L-1} \delta _j ^2 +O(\delta _n ^5) =  \frac {\pi}{2} k_n \, ,
\end{equation}
which we rewrite as
\begin{equation}
\delta _n= \frac {\frac {\pi}{2} k_n}{2 \ln s-L \psi \left (\frac {1}{2}\right ) +\frac {L}{6} \psi ^{(2)}\left (\frac {1}{2}\right ) \delta _n ^2 +(L-2) \psi (1) - \frac {L-2}{6}\psi ^{(2)}(1) \delta _n ^2 - \frac {1}{2}\psi ^{(2)}(1) \sum _{j=2}^{L-1} \delta _j ^2 +O(\delta _n ^4)} \, .
\end{equation}
Therefore, we can write that
\begin{eqnarray}
\delta _n&=& \frac {\frac {\pi}{2} k_n}{2 \ln s-L \psi \left (\frac {1}{2}\right )  +(L-2) \psi (1)}\Bigl [ 1-\frac {\frac {L}{6} \psi ^{(2)}\left (\frac {1}{2}\right ) \delta _n ^2 - \frac {L-2}{6}\psi ^{(2)}(1) \delta _n ^2 - \frac {1}{2}\psi ^{2}(1) \sum _{j=2}^{L-1} \delta _j ^2 +O(\delta _n ^4)}{2 \ln s-L \psi \left (\frac {1}{2}\right )  +(L-2) \psi (1)}+ \nonumber \\
&+& O\left (\frac {\delta _n ^4}{(\ln s)^2}\right ) \Bigr ]\, .
\end{eqnarray}
>From (3.37) of \cite {BGK} we get that in the large $s$ limit the
energy $E_0$ depends on $\delta _n$ according to the formula
\begin{equation}
E_0=4\ln s - 2L \psi (1)+ \sum _{n=2}^{L-1}\left [ \psi \left (\frac {1}{2}+i\delta _n \right )+\psi \left (\frac {1}{2}-i\delta _n \right ) \right ] \, .
\end{equation}
Developing for small $\delta _n$ we have
\begin{equation}
E_0=4\ln s-2L\psi (1)+\sum _{n=2}^{L-1}\left [ 2 \psi \left (\frac {1}{2} \right )- \psi ^{(2)} \left (\frac {1}{2} \right )\delta _n ^2 +\frac {1}{12} \psi ^{(4)} \left (\frac {1}{2} \right )\delta _n ^4 + O(\delta _n^6) \right ] \, .
\label{E0bgk}
\end{equation}
Now, for the ground state $k_n=L+1-2n$. Since we want to compare this formula
with (\ref {E0}), we have to consider only terms in the energy which go as
$\ln s \ j^k$ and $j^k$, with $1\leq k \leq 5$.
In such an approximation the relevant sums appearing in (\ref {E0bgk}) read as
\begin{eqnarray}
\sum _{n=2}^{L-1} \delta _n ^2&=&\frac {\frac {\pi ^2}{4} }{[2 \ln s-L \psi \left (\frac {1}{2}\right )  +(L-2) \psi (1)]^2 } \cdot \\
&\cdot &  \sum _{n=2}^{L-1} \Bigl [ k_n^2-2 \frac {\frac {L}{6} \psi ^{(2)}\left (\frac {1}{2}\right ) \frac {\pi^2}{16 (\ln s)^2} k_n^4 - \frac {L-2}{6}\psi ^{(2)}(1)
\frac {\pi^2}{16 (\ln s)^2} k_n^4 - \frac {1}{2}\psi ^{2}(1) \frac {\pi^2}{16 (\ln s)^2} k_n^2 \sum _{j=2}^{L-1} k_j^2 }{2 \ln s-L \psi \left (\frac {1}{2}\right )  +(L-2) \psi (1)} +\ldots \Bigr ] \, , \nonumber \\
\sum _{n=2}^{L-1} \delta _n ^4&=&\frac {\frac {\pi ^4}{16} }{[2 \ln s-L \psi \left (\frac {1}{2}\right )  +(L-2) \psi (1)+\ldots ]^4}  \sum _{n=2}^{L-1} k_n^4
\end{eqnarray}
Now, using the sums
\begin{equation}
\sum _{n=2}^{L-1} k_n^2= \frac {(L-2)^3}{3} + O(L-2) \, , \quad
\sum _{n=2}^{L-1} k_n^4= \frac {(L-2)^5}{5} + O(L-2)^3 \, ,
\end{equation}
and also the fact that
\begin{equation}
2 \ln s-L \psi \left (\frac {1}{2}\right )  +(L-2) \psi (1)=2\ln s \left [1+\frac {2\ln 2 +\gamma _E}{\ln s}+ \ln 2 \ j \right ]  \, ,
\end{equation}
one gets, at the order of interest,
\begin{eqnarray}
\sum _{n=2}^{L-1} \delta _n ^2&=&\frac {\pi ^2 (L-2)^3}{48 (\ln s)^2 \left (1+
\frac {2\ln 2 +\gamma _E}{\ln s} \right )^2 } \left [1-\frac {2\ln 2 }{\left
(1+ \frac {2\ln 2 +\gamma _E}{\ln s} \right )}j+\frac {3 (\ln 2)^2 }{\left
(1+ \frac {2\ln 2 +\gamma _E}{\ln s} \right )^2} j^2 +\ldots \right ] - \nonumber \\
&-& \frac {\pi ^4}{3840}\frac {\psi ^{(2)}\left (\frac {1}{2}\right ) (L-2)^5}
{(\ln s)^5} \, . \\
\sum _{n=2}^{L-1} \delta _n ^4&=&\frac {\pi ^4}{1280}\frac {\ln s }
{\left (1+ \frac {2\ln 2 +\gamma _E}{\ln s} \right )^4 }j^5 \, ,
\end{eqnarray}
where, in order to keep formul{\ae} compact, we did not develop in inverse powers of $\ln s $ the brackets in the denominators.
With the help of the explicit values
\begin{equation}
\psi ^{(2)}\left (\frac {1}{2}\right )=-14 \zeta (3) \, , \quad
\psi ^{(4)}\left (\frac {1}{2}\right )=-744 \zeta (5) \, ,
\end{equation}
we eventually plug all the results in (\ref {E0bgk}), getting the expression
\begin{eqnarray}
E_0&=& 4 \ln s  - 4 \ln s \ \ln 2 \ j + 4\gamma _E +
\ln s \frac {7 \zeta (3) \pi ^2}{24}  \frac {j^3}{\left
(1+ \frac {2\ln 2 +\gamma _E}{\ln s} \right )^2} -
\ln s \frac {7 \zeta (3) \pi ^2 \ln 2}{12} \frac {j^4}{\left
(1+ \frac {2\ln 2 +\gamma _E}{\ln s} \right )^3} + \nonumber \\
&+& \ln s \left [\frac{7\pi^2 (\ln 2)^2 \zeta(3)}{8}-\frac{31\pi^4\zeta(5)}{640}\right ] \frac {j^5}{\left
(1+ \frac {2\ln 2 +\gamma _E}{\ln s} \right )^4} +
\frac{49\pi^4 \zeta^2 (3)}{960} j^5 + \ldots
\label {E0bgk2} \, .
\end{eqnarray}
Finally, developing the brackets in the denominators in their $\ln s$ and $(\ln s)^0$ contribution,
we get formula (\ref {E0}).

\end{document}